\documentclass[12pt]{article}
\usepackage[polish,english]{babel}
\usepackage[latin1]{inputenc}
\usepackage{times}
\usepackage[T1]{fontenc}
\usepackage{epsfig}

\begin{document}

\title{Swaying oscillons \\ in the signum-Gordon model }

\author{H. Arod\'z$^a$ and $\;$ Z. \'Swierczy\'nski$^b$ \\$\;\;$ \\ \emph{\small $^a$ Institute of Physics,
Jagiellonian University, Cracow, Poland }\\ \emph{\small
$^b$Institute of Computer Science and Computer Methods,}\\
\emph{\small Pedagogical University, Cracow, Poland}}

\date{$\;$}

\maketitle

\begin{abstract}
We present a new class of oscillons in the (1+1)-dimensional
signum-Gordon model. The oscillons periodically move to and fro in
the space.  They have finite total energy, finite size, and are
strictly periodic in time.  The corresponding solutions of the
scalar field equation are explicitly constructed from the second
order polynomials in the time and position coordinates.
\end{abstract}

\vspace*{2cm} \noindent PACS: 05.45.-a, 03.50.Kk, 11.10.Lm \\

\pagebreak

\section{ Introduction}

Scalar fields play essential role in many branches of physics, from
cosmology to condensed matter physics to particle physics --  there
is an unremitting interest in models of self-interacting scalar
fields.  The rich variety of such models includes some that have
been studied only recently, e.g., the so called K-fields with a
nonstandard kinetic part \cite{1}, or models with a non-smooth
V-shaped self-interaction  \cite{2}. The signum-Gordon model
considered in the present paper is probably the simplest example
from the latter class. The pertinent field potential has the form
$U(\varphi)=  g |\varphi|,$ where $g
>0$ is a coupling constant and $|\varphi|$ is the modulus of the real
scalar field $\varphi$. Such potential is V-shaped with the sharp
minimum at the vacuum field $\varphi=0$. Models of this kind were
discovered while playing with the well-known classical systems of
harmonically coupled pendulums in order to illustrate the phenomenon
of spontaneous symmetry breaking and topological defects \cite{3}.
Subsequent investigations have revealed that the V-shaped form of
the potential has very interesting consequences for the dynamics of
the scalar field. One of them is the existence of strictly periodic
oscillons \cite{4}. The motivation, various results and further
references for the V-shaped self-interaction can be found in
\cite{2,3,4}.

The present note is a follow-up to the paper \cite{4}. The oscillons
described in that paper did not move in space (apart from the
trivial uniform motion obtained by applying Lorentz boosts). Rather
unexpectedly, we have found that there exist also oscillons that
periodically move to and fro in the space with arbitrary constant
velocity $\pm v$, where $0 < |v| \leq 1$. For the oscillons
presented in \cite{4}  $v=0$. The  new oscillons appear naturally
when the particular solution reported in \cite{4} is put in the
framework of polynomial solutions of the signum-Gordon equation.

Comparing with other oscillons discussed in literature \cite{5},
several differences should be pointed out. First, our oscillons are
strictly periodic in time, in particular they do not emit any
radiation. Second, they have  strictly finite size because the field
assumes the vacuum value  at a finite distance exactly. Third, they
have relatively simple, explicitly given form composed of several
linear and quadratic functions of the time $t$ and the spatial
coordinate $x$.

The swaying oscillon reminds  the wobbling kink in the $\varphi^4$
model \cite{6}.  However, one should note that the wobbling kink is
an excitation of static kink, while all the swaying oscillons are
degenerate in energy, and moreover there is no static oscillon --
even for the presented in \cite{4} non-swaying one  the field
oscillates in time.

The plan of our paper is as follows. Section 2 is devoted to a
preliminary discussion of the signum-Gordon equation and of its
solutions. The swaying oscillons are presented in Section 3. Section
4 contains the conclusion.

\section{Preliminaries }

The Lagrangian of the signum-Gordon model (s-G) has the form
\begin{equation} L = \frac{1}{2} (\partial_t \varphi\partial_t
\varphi  -   \partial_x \varphi\partial_x \varphi)
 - g\:|\varphi|, \end{equation} where $\varphi$ is a real  scalar field,
$t, x$ are  time and position coordinates in the two-dimensional
Minkowski space-time $M$. For convenience, $t, x, \varphi, g$ are
dimensionless --  this can  be achieved by redefinitions of the
physical position, time, field and the coupling constant
(multiplication by constants of appropriate dimensions).  The
signum-Gordon equation
\begin{equation}
\partial_t^2 \varphi     -  \partial_x^2 \varphi + \mbox{sign}(\varphi(x,
t))=0
\end{equation}
is the Euler-Lagrange equation corresponding to Lagrangian (1) (from
now on we put $g=1$). The sign function $\mbox{sign}(\varphi)$ has
the values $\pm 1$ for $\varphi \neq 0$ and $\mbox{sign}(0)=0$. The
simplest way to obtain Eq.\ (2) from Lagrangian (1) is first to
regularize the field potential $U(\varphi)= |\varphi|$, e.g.,
$U(\varphi) =  \sqrt{\epsilon^2 + \varphi^2}$ or $U(\varphi) =
\epsilon\: \ln(\cosh(\varphi/\epsilon))$, and to take the limit
$\epsilon \rightarrow 0_+ $ in the Euler-Lagrange equation obtained
from the regularized Lagrangian. Direct computation of the variation
of the action $S = \int dt dx L$ is more subtle because of the
$|\varphi|$ term, but  it  gives the signum-Gordon equation (2) too.

The l.h.s.  of  Eq.\ (2) is not continuous with respect to
$\varphi$. Because such equations are not very common in field
theory, let us briefly comment  on the related mathematical aspects.
First, it is clear that in general one should expect non smooth
solutions: the value of at least one of the second derivatives
$\partial_t^2\varphi, \partial_x^2\varphi$ has to jump when the
function $\mbox{sign}(\varphi)$  changes its value. Second, the use
of the stationary action principle implies that in general we
consider so called weak solutions of the Euler-Lagrange equation,
\cite{7}. For the weak ones it is sufficient that
\[ \delta S = \int_M dt dx\; \left(\frac{\partial L}{\partial
\phi}\:
\delta\phi(x,t)
+ \frac{\partial L}{\partial (\partial_{\mu}\phi)}  \;
\partial_{\mu} \delta\phi(x,t)\right) =0 \] for all test functions $\delta\phi(x,t)$ from a
certain class (typically one uses the  $D(M)$ class of smooth
functions on $M$ with compact support).  This condition is
equivalent to $\int_M dt dx\:  {\cal E\!\!L}\: \delta \varphi =0$,
where $ {\cal E\!\!L} =
\partial L/\partial \phi - \partial_{\mu}(\partial L/
\partial(\partial_{\mu}\varphi))$, only if the derivative $\partial_{\mu}(\partial L/
\partial(\partial_{\mu}\varphi))$ exists for a given probed function
$\varphi(x,t)$. Then  the Euler--Lagrange equation ${\cal
E\!\!L}=0$, in our case the signum-Gordon Eq.\ (2), has to be
satisfied at almost all points $(x,t)$ in the two-dimensional
space-time  $M$, but not necessarily at all points as it would be
the case with strong solutions.  Of course, the set of weak
solutions contains  the strong ones as a subset.

In the case of signum-Gordon equation the weak solutions that are
not strong are ubiquitous.  For instance, $\varphi_0 = x^2/2$ is a
smooth static solution of Eq.\ (2) in the weak sense, but not in the
strong sense. The point is that $\partial_t^2 \varphi_0 -
\partial_x^2 \varphi_0 + \mbox{sign}(\varphi_0)=0$ everywhere in
$M$ except the line $x=0$ in $M$. On this line $\partial_t^2
\varphi_0 -
\partial_x^2 \varphi_0 + \mbox{sign}(\varphi_0)= -1 $
because $\partial_x^2 \varphi_0 =1$, $ \mbox{sign}(0) =0$.
Nevertheless,
\[ \int_{M} dt dx \;[\partial_t^2 \varphi_0 -
\partial_x^2 \varphi_0 + \mbox{sign}(\varphi_0)] \; \delta\phi(x,t)
=0 \] for arbitrary test function $\delta\phi$.

In general, physically relevant are the weak solutions.  To see
this, consider the following simple example from classical mechanics
of a point particle on a plane with Cartesian coordinates  $(x,y)$.
The particle is free except when it crosses the $y$-axis, where  it
is subjected to a finite constant force $\vec{F}_0$ parallel to the
$y$-axis. Thus, the force $\vec{F}=0$ at all points $(x, y)$ with
$x\neq0$,  and $\vec{F}= \vec{F}_0$ when  $x =0$. It is clear that
integrating the Newton's equation $d\vec{p}/dt = \vec{F}$ we obtain
$\vec{p} = const$ even if the trajectory crosses the $y$-axis.  The
physical reason is that the finite force $\vec{F}_0$ acts on the
particle only during infinitesimally short time when the particle is
exactly on the $y$-axis, hence it is not able to perturb the free
motion. Such trajectories are the weak solutions of the Newton's
equation (now the test functions are denoted as $\delta\vec{r}(t)$
and we integrate over $t$).  On the other hand, the trajectories
which do not intersect  the $y$-axis are solutions in the strong
sense. Notice that such Newton's equation is not equivalent to the
free equation, in which $\vec{F}=0$ everywhere,  because our
particle is accelerated if it moves along the $y$-axis.

Coming back to the signum-Gordon model, in the case  the field
$\varphi$ is constant in the space  Eq. (2) acquires the form of one
dimensional Newton's equation $ \ddot{\varphi}(t) = -
\mbox{sign}(\varphi)$ that describes nonlinear oscillations around
$\varphi=0$. Notice that there is no linear regime even for
arbitrarily small values of $\varphi$. Newton's equation of this
kind appears in the elementary problem of a ball vertically bouncing
from a floor in a constant gravitational field (the elevation above
the floor is given by $|\varphi|$) \cite{2}.   Many examples of
oscillatory systems from classical mechanics that do not have the
linear small amplitude regime can be found in \cite{8}.

Because the function $\mbox{sign}(\varphi)$ is piece-wise constant,
it is natural first to solve Eq.\ (2) in the regions in which
$\varphi$ has a constant sign. For instance, if $\varphi <0$, Eq.\
(2) acquires the form
\begin{equation}
\partial_t^2 \varphi     -  \partial_x^2 \varphi - 1 =0.
\end{equation}
The oscillon solutions are constructed from second order polynomials
in $x, t$.  The most general second order polynomial  that obeys Eq.
(3) has the form
\begin{equation}
\varphi_2(x, t) = a_0 x^2 + a_1 t x + (a_0+\frac{1}{2}) t^2 + b_0 x
+ b_1 t + c_0,
\end{equation}
where $a_0, a_1, b_0, b_1, c_0$ are constants  (beware that they are
not completely arbitrary because of the condition $\varphi_2 <0$).
It is rather exceptional feature of the signum-Gordon equation that
non-trivial and interesting solutions can be constructed from such
simple elementary functions.  Note that the class of functions of
the form (4) is invariant with respect to Lorentz boosts, space-time
translations, and the reflections  $x \rightarrow -x, \; t
\rightarrow -t$. It contains the static solutions of the form
\begin{equation}
\varphi_s = - \frac{1}{2} (x-b_0)^2 + c_0 + \frac{1}{2} b_0^2,
\end{equation}
where $c_0 + b_0^2/2 <0$ in order to keep $\varphi_s <0$.

The oscillons are constructed by patching together several such
polynomial solutions. The patching conditions have the standard
form: the field $\varphi$ is continuous all over  $M$, also the
derivatives $\partial_t\varphi, \partial_x\varphi$ are continuous
function of $x, t$ except perhaps at the border line between two
patches. If the border line  is  a (segment of)   characteristic
line ($x = \pm t + \mbox{const}$) for the signum-Gordon equation,
the  derivative in the direction perpendicular to that line does not
have to be continuous --  a finite jump is allowed.

\section{The swaying oscillons }

Hint that new oscillons  may exist comes from the following
procedure for constructing periodic solutions of the signum-Gordon
equation. Let $\varphi_-(x,t) $ be a  solution of Eq.\ (3) negative
for all $t$ from an open interval  $(0, T),  \; T>0,$ and such that
\begin{equation}
\varphi_-(x, 0) = \: 0 \: = \varphi_-(x, T).
\end{equation}
It is clear that the function $\varphi_+$ defined by
\begin{equation}
\varphi_+(x, t) = -  \varphi_-(x, -t)
\end{equation}
is a positive solution of the equation $\partial_t^2 \varphi     -
\partial_x^2 \varphi +1 =0   $ for all $t \in (-T,0)$.
The functions $\varphi_-, \varphi_+$ as well as their time
derivatives match each other at the time $t=0$:
\[
\varphi_+(x, 0) = \: 0 \: = \varphi_-(x, 0), \]
\[
\lim_{t \rightarrow 0_-} \partial_t\varphi_+(x, t ) = - \lim_{t
\rightarrow 0_-} \partial_t\varphi_-(x, -t )  = \lim_{s \rightarrow
0_+} \partial_s\varphi_-(x, s ),
\]
where $s$ stands for $-t$, and $t\in (-T,0)$.  The crucial
observation is that also $\varphi_+(x, -T), \:  \varphi_-(x, T)$
match each other:
\[
\varphi_+(x, -T) = \: 0 \: = \varphi_-(x, T), \]
\[
\lim_{t \rightarrow -T_+} \partial_t\varphi_+(x, t ) =  \lim_{s
\rightarrow T_-} \partial_s\varphi_-(x, s ).
\]
Therefore, we may extend our partial solutions $\varphi_{\pm}$ to
all times $t \geq T$ and $ t \leq -T$ just by applying time
translations (by multiples of $\pm T$) to $\varphi_{\pm}$. In this
way we obtain periodic solutions of the signum-Gordon equation  (2)
with the period equal to $2T$, provided that there exists
$\varphi_-(x,t)$ with the properties specified above.  It turns out
that a class of the solutions  $\varphi_-(x,t)$ with the desired
properties can be constructed by patching together several solutions
of the form (4). Also the trivial solution $\varphi=0$ is involved.
The schematic picture of such `patchwork' for the swaying oscillon
is presented in Fig.\ 1.

Note that so far we have not made any assumption about the behavior
of $\varphi_-$ at large $|x|$. In \cite{4} certain restrictive
boundary conditions were imposed right at the start of calculations
and they forced the oscillons to stay still.  The method adopted in
the present note is radically different from the one used in
\cite{4} -- in that paper the main tool was d'Alembert formula for
solutions with given initial data.

In order to ensure finiteness of the total energy we assume that
$\varphi_-(x,t) =0$ outside a certain compact region. This is
related to the general observation that in the case of models with
the V-shaped potential there are no exponential or long range tails.
The field reaches its vacuum value rather abruptly, the tails have a
parabolic shape and a strictly finite length \cite{2}. Thus, our
first task is to find the polynomials of the form (4) which match
the trivial solution $\varphi =0$. The matching conditions imposed
on a line $x(t)$ in $M$ can be written in the form
\[
\varphi_2(x(t), t) =0, \;\;\; \left. \partial_x \varphi_2(x,
t)\right|_{x = x(t)} =0. \]  They give the following two equations
\[
a_0 x^2(t) + a_1\: t\: x(t)  + (a_0+\frac{1}{2}) t^2 + b_0 x(t) +
b_1 t + c_0 =0, \;\;\; 2 a_0 x(t) + a_1 t + b_0 =0.  \] Simple
calculations show that the solution $x(t)$ exists only if $a_0
\neq0$, and then
\begin{equation}
x(t) = v t + x_0, \;\; \varphi_2(x,t) = -\frac{(x - x(t))^2}{2
(1-v^2)},
\end{equation} where  $ v= - a_1/(2 a_0), \; x_0 = - b_0/(2 a_0)$,
and $v^2 <1$  in order to satisfy the condition $\varphi_2 < 0$.
Here we have assumed that $x(t)$ does not coincide with a
characteristic line.

Thus we have found that the boundary of our oscillon has to move
with the constant velocity $v$, and close to the boundary the field
has the parabolic shape (as expected). Note that $\varphi$ given by
formula (8) coincides with the Lorentz boosted and translated in the
space the static solution
\[
\varphi_s = \left\{ \begin{array}{cc} 0  &  x \leq 0, \\
- \frac{x^2}{2} &  x>0. \end{array}\right.
\]
(but the swaying oscillon is not the Lorentz boosted oscillon of
\cite{4}).

The structure of the solution $\varphi_-$ is shown in Fig.\ 1, in
which the support of $\varphi_-$ for the swaying oscillon of unit
length and vanishing total momentum is depicted. The period of this
oscillon is equal to its spatial size, i.e., to 1. As discussed in
\cite{4}, we may use the symmetries of the signum-Gordon equation,
such as Poincar\'e or scaling transformations in order to obtain
more general oscillons. The interior of the parallelogram is divided
into seven sectors $\emph{a}\div \emph{f}\:$ by the four
characteristic lines drawn from its corners. Each sector has
different causal neighborhood. For instance, the field in the
triangular sector \emph{c} is completely determined by Cauchy data
on the segment $[1/2 + v/2, 1]$ of the $x$ axis; the sector \emph{e}
is controlled by Cauchy data in the future, i.e., on the segment
$[1/2, 1+v/2]$ of the $t=1/2$ line which lies in the future of the
sector \emph{e}; the sectors \emph{a } and \emph{d} are controlled
by the boundaries of the oscillon, etc. The parallelogram shown in
Fig.\ 1 has the height equal to one half of its length. In this case
the characteristic lines drawn from the lower (upper) corners meet
at a point lying on the upper (lower) edge.

In the case of the non-swaying oscillon presented in \cite{4} we
have $v=0$ and a rectangle in Fig.\ 1. The parallelogram is the
simplest deformation of that rectangle  consistent with the
conditions $ \varphi_-(x,0) = 0 = \varphi_-(x, \frac{1}{2})$,  and
with the fact that the both sides of the oscillon have to move with
a constant velocity, as has been shown above. Such generalization --
the parallelogram instead of the rectangle -- is very suggestive in
the `patchwork approach' adopted in the present paper, but it is not
obvious at all in the based on the d'Alembert formula approach used
in \cite{4}. Let us also note that a Lorentz boost of the oscillon
considered in \cite{4} gives a uniform rectilinear motion, and not
the swaying one.  Moreover, it deforms the rectangle into a
hyperbolically rotated parallelogram with the upper and bottom sides
not parallel to the $x$-axis.

The building blocks of $\varphi_-$ are denoted as $\varphi_a, \ldots
, \varphi_g$ after the sectors of the parallelogram. The fields
$\varphi_a, \varphi_d$ in the sectors $a$ and $ d$  have the form
(8) with $x(t) = v t$ or $x(t)= v t +1$, respectively, i.e.,
\begin{equation}
\varphi_a = -\frac{(x-vt)^2}{2(1-v^2)}, \;\;\; \varphi_d = -
\frac{(x-vt-1)^2}{2(1-v^2)}.
\end{equation}
In the regions $x \leq vt$ and $ x \geq v t +1$, i.e. on both sides
of the parallelogram, the field has the vacuum value  $\varphi=0$.
\begin{center}
\begin{figure}[tph!]
 \hspace*{1.7cm}
\includegraphics[height=5.5cm, width=8cm]{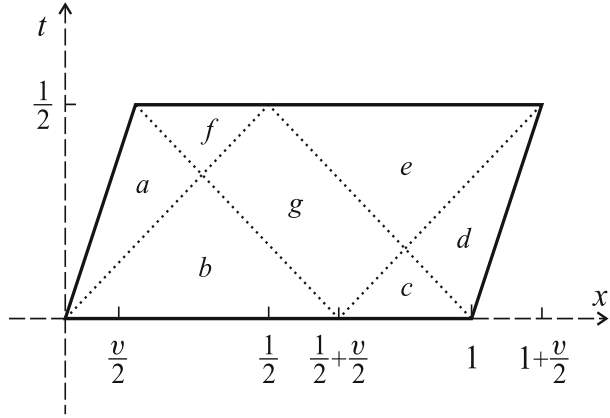}
\caption{\small The support of the  solution $\varphi_-(x,t)$. The
field  $\varphi_-(x,t)$ vanishes on the continuous lines that form
the boundary of the parallelogram. In each sector $\emph{a}\div
\emph{f}$ the function $\varphi_-$ is given by a different formula.
The matching conditions that relate the functions in neighboring
sectors are imposed along the dotted lines. These four lines  are
characteristic lines of the signum-Gordon equation. They have the
slopes $\pm 1$. }
\end{figure}
\end{center}
The fields $\varphi_b, \varphi_c,  \varphi_e,  \varphi_f$ are
determined by imposing on the  solution (4) the condition $\varphi_2
=0$ on the lines $t=0$ or $t=1/2$, and  the  conditions of matching
with $\varphi_a$ or $\varphi_d$ on the characteristic lines. As the
example let us determine $\varphi_b$. The condition $\varphi_2(x,0)
=0$ gives $a_0=b_0=c_0=0$. Next, $\varphi_2 = t^2/2 + a_1 t x +b_1
t$ is compared to $\varphi_a$ on the part of the characteristic line
$x=t$ with $t \in (0, 1/4 + v/4)$:
\[ \frac{t^2}{2} + a_1 t^2 +b_1 t = - \frac{1-v}{2(1+v)} t^2. \] Therefore,
$b_1=0$, $a_1 = -1/(1+v)$, and  $\varphi_b = t^2/2 - tx/(1+v)$.
Similar calculations give $\varphi_c, \varphi_e, \varphi_f$.
Finally, we compute $\varphi_g$ by comparing $\varphi_2$ to
$\varphi_b, \varphi_c,  \varphi_e, \varphi_f$  along the four
characteristic lines that form the boundary of the sector $g$. The
results have the following form:
\begin{equation}
\varphi_b = \frac{t^2}{2} - \frac{t x }{1+v}, \;\;\; \varphi_c =
\frac{t^2}{2} + \frac{t (x-1)}{1-v},
\end{equation}
\begin{equation}
\varphi_e = \frac{1}{2} (t-\frac{1}{2}) \: \left( \frac{1 }{2 } + t
+ \frac{1-2x }{1+v}\right), \end{equation}
\begin{equation}
\varphi_f = \frac{1}{2} (t-\frac{1}{2}) \: \left( \frac{1 }{2 } + t
+ \frac{2x-1}{1-v}\right),
\end{equation}
\begin{equation}
\varphi_g =  \frac{(v x +t)^2 }{2(1-v^2)} + \frac{x^2 + t^2}{2 } +
\frac{1+v}{8(1-v)}  - \frac{x+t }{2(1-v)}.
\end{equation}
All these functions are negative inside their domains.  The shape of
the swaying oscillon is depicted in Fig.\ 2.
\begin{center}
\begin{figure}[tph!]
\hspace*{1.8cm}
\includegraphics[height=5.5cm, width=7cm]{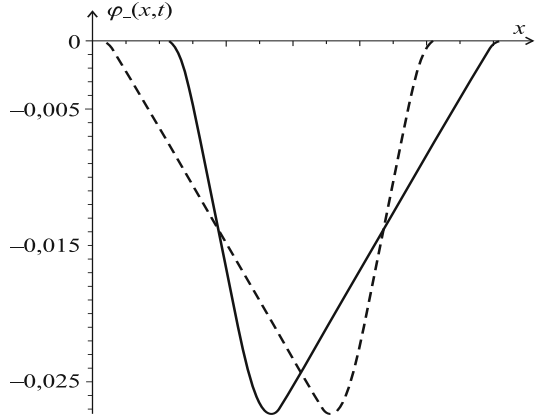}
\caption{\small The shape of the swaying oscillon at the times
$t=1/8$ (the dashed line) and $t=3/8$ (the continuous line). The
velocity of the swaying motion  $v = 1/2$. }
\end{figure}
\end{center}
The evolution of our oscillon is described by the function
$\varphi_-(x, t)$ in the time interval $[0, 1/2]$, and by
$\varphi_+(x,t)$,  formula (7),  for $t\in[-1/2,0]$. In particular,
the field $\varphi_+$ at the boundaries of the oscillon has the form
\[
\varphi_{+,a}(x,t) = \frac{(x+v t)^2}{2(1-v^2)}, \;\;\;
\varphi_{+,d}(x,t) = \frac{(x+v t-1)^2}{2(1-v^2)}.  \] We see that
now the boundaries of the oscillon move with  the velocity $-v$. The
world-sheet of the oscillon is depicted in Fig.\ 3. Note that at the
times $t=k/2$, $k$ - integer, when the sharp turns take place, the
field $\varphi$ vanishes everywhere.
\begin{center}
\begin{figure}[tph!]
 \hspace*{1.8cm}
\includegraphics[height=5.5cm, width=7cm]{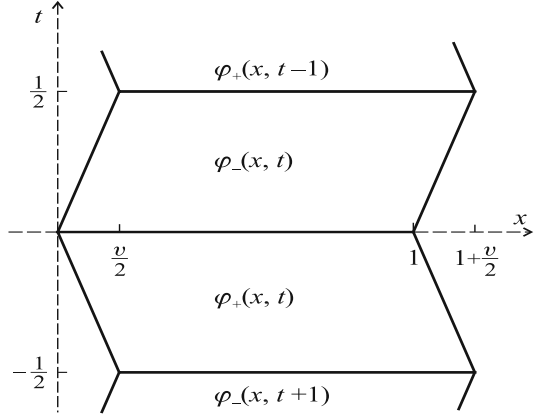}
\caption{\small The world-sheet of the swaying oscillon.  In the
interior of the parallelograms $\varphi_+ >0$ and $\varphi_- <0$,
whereas on their boundaries (the thick continuous lines)
$\varphi_{\pm}=0$.}
\end{figure}
\end{center}

In the case  $x(t)$ is a  characteristic line we have $x(t) = v t +
x_0$, where  $|v| =1$. There is just one matching condition
$\varphi_2(x(t),t) =0$. Solving it we obtain relations between the
constant coefficients present in $\varphi_2$, and finally
\[
\varphi_2(x,t) = (x-x(t))\; \left(a_0 \:(x - x(t)) - \frac{1}{2}
x(t) + (\frac{1}{2} + 2 a_0) x_0 + b_0\right) \] (in the region
where $\varphi_2(x,t) <0$).  Next steps are similar to those
described above, but the situation is much simpler. In particular,
when $v=1$, the left and right hand sides of the parallelogram in
Fig.\ 1 coincide with characteristic lines. Therefore, the sectors
$a, f, c, d, g$ are absent. The remaining sectors $ b, e$ meet at
the line $x=1-t$. The corresponding functions $\varphi_b, \varphi_e$
are given by formulas (10), (11) with $v=1$, and they correctly
match each other on that line.

The total energy $E$ and momentum $P$ of the oscillon can easily be
calculated from formulas
\[
E= \frac{1}{2}  \int^{\infty}_{-\infty}dx\; [(\partial_t\varphi)^2 +
(\partial_x\varphi)^2]+  \int^{\infty}_{-\infty}dx\; |\varphi|,
\;\;\; P = -  \int^{\infty}_{-\infty}dx\;\partial_t\varphi
\partial_x\varphi,  \]
considered at the time $t=0$ when $\varphi=0= \partial_x\varphi$. We
see that $P=0$, in spite of the swaying motion of the oscillon. This
can be understood if we regard the swaying oscillon as a nonlinear
bound state of the basic oscillon, that is the one with  $v=0$, with
a wave packet traveling along the basic oscillon. If the swaying
oscillon has  $P=0$, the nonzero momentum of the wave packet is
compensated by the momentum related with the motion of the basic
oscillon. The wave packet bounces from the boundaries of the basic
oscillon and does not leave its interior. Then the basic oscillon
has to move accordingly in order to keep $P=0$.

In order to compute the total energy we need
$\partial_t\varphi|_{t=0}$. Formulas (9), (10)  give
$\partial_t\varphi_b|_{t=0}= -x/(1+v)$ for $ x\in [0,(1+v)/2]$, and
$\partial_t\varphi_c|_{t=0}= (x-1)/(1-v)$ for $ x\in [(1+v)/2, 1]$.
In the case $v=\pm1$ the part with $\varphi_c$ is absent.  Simple
integration gives $E=1/24$. Thus the total energy does not depend on
$v$ --  all our swaying oscillons have the same energy. We have not
found any explanation for such a degeneracy. One may suspect that
there exists a hidden symmetry. Note that it would be sufficient if
it works only in the subspace of the polynomial solutions
$\varphi_2$, not necessarily on the level of Lagrangian or action.
The bound state interpretation offers the following picture. The
basic oscillon set in motion would have an energy larger that its
rest energy equal to 1/24. Apparently, the binding energy
compensates the kinetic energy of the basic oscillon as well as the
energy of the bouncing wave packet, so that the total energy remains
equal to 1/24.

\section{Conclusion}

 We have shown that oscillons in   the (1+1)-dimensional
signum-Gordon model can periodically move to and fro in the space
(the x-line) with a constant speed $v$ from the interval $[0,1]$.
The amplitude of such swaying motion is equal to $v l/2$, where $l$
is the length of the oscillon. The pertinent analytic solutions of
the field equation have been constructed from the second order
polynomials in $t$ and $x$.

The present paper is a follow-up to \cite{4}, and the remarks and
comments given there apply also to the swaying oscillons. Our new
findings contribute to the already substantial evidence that the
models  of the signum-Gordon type have rather amazing properties. In
particular, it is quite surprising that one can find simple,
explicit solutions that describe very nontrivial objects like the
oscillons, or $Q$-balls \cite{9}, and this happens in spite of the
unpleasant $\mbox{sign}(\varphi)$ form of the nonlinear term in the
field equation.


\begin{thebibliography}{99}
\bibitem{1} C. Armendariz-Picon, T. Damour and V. F. Mukhanov, Phys.
Lett. B \textbf{458}, 2009 (1999); C. Adam, P. Klimas,  J.
Sanchez-Guillen and A. Wereszczy\'nski, J. Math. Phys. \textbf{50}:
102303 (2009).
\bibitem{2} H. Arod\'z, P. Klimas and T. Tyranowski, Acta Phys. Pol. B \textbf{36}, 3861
(2005); Phys. Rev. E \textbf{ 73}, 046609 (2006).
\bibitem{3} H. Arod\'z, Acta Phys. Pol. B \textbf{33}, 1241 (2002);
\emph{ibidem} \textbf{35}, 625 (2004).
\bibitem{4}  H. Arod\'z, P. Klimas and T. Tyranowski,  Phys. Rev. D
\textbf{77}, 047701 (2008).
\bibitem{5} See, e.g., I. L. Bogolyubskii and V. G. Makhankov, JETP Lett. \textbf{24}, 12 (1976);
 M. Gleiser, Phys. Lett. B \textbf{600}, 126 (2004); M. Hindmarsh and P. Salmi, Phys. Rev.
D \textbf{74}, 105005 (2006); G. Fodor, P. Forgacs, P.
Grandcl\'ement and I. R\'acz, Phys. Rev. D \textbf{74}, 124003
(2006);  M. Gleiser and J. Thorarinson, Phys. Rev. D \textbf{76},
041701(R) (2007); M. Gleiser and D. Sicilia, Phys. Rev. Lett.
\textbf{101}, 011602 (2008).
\bibitem{6} See, e.g.,  I. V.  Barashenkov  and  O. F. Oxtoby, Phys. Rev. E \textbf{80}, 026608 (2009);  O. F. Oxtoby and I. V.  Barashenkov, Theor. Math. Phys.
\textbf{159}, 863 (2009).
\bibitem{7} See. e.g.,  R. D. Richtmyer, \emph{Principles of Advanced Mathematical Physics}. Springer-Verlag,
New York-Heidelberg-Berlin, 1978. Section 17.3.  L. C. Evans,
\emph{Partial Differential Equations.} American Math. Society, 1998.
\bibitem{8}  P. Mohazzabi, Am. J. Phys. \textbf{72}, 492 (2004).
\bibitem{9} H.  Arod\'z  and J. Lis,   Phys. Rev. D \textbf{77}, 107702 (2008).
\end{thebibliography}
\end{document}